# An electrically pumped phonon-polariton laser

Keita Ohtani, Bo Meng, Martin Franckié*, Lorenzo Bosco, Camille Ndebeka-Bandou, Mattias Beck, Jérôme Faist

**Affiliations:**

Institute for Quantum Electronics, ETH Zurich, August-Piccard-Hof 1, 8093 Zurich, Switzerland

*Correspondence to: martin.franckie@phys.ethz.ch

**Abstract:** We report a device that provides coherent emission of phonon polaritons, a mixed state between photons and optical phonons in an ionic crystal. An electrically pumped GaInAs/AlInAs quantum cascade structure provides intersubband gain into the polariton mode at λ = 26.3 μm, allowing self-oscillations close to the longitudinal optical phonon energy of AlAs. Because of the large computed phonon fraction of the polariton of 65%, the emission appears directly on a Raman spectrum measurement exhibiting a Stokes and anti-Stokes component with the expected shift of 48 meV.

**One Sentence Summary:** We report the direct observations of coherently emitted phonon polaritons via their photon, phonon, and polariton signatures.

**Main Text:** The polariton, a mixed state between a photon and an electronic excitation in solid-state matter, has recently attracted much attention because of its rich physics that includes superfluidity, quantized vortices and Bose-Einstein condensation (*1*, *2*). In addition, the possibility of engineering both the optical and matter part of this quasi-particle can be exploited for the creation of polaritonic devices with enhanced non-linear properties (*3*-*5*). An exciton-polariton laser created using a semiconductor microcavity is an elegant example since it reaches threshold by stimulated scattering of the polaritons (*6*, *7*) instead of population inversion between conduction and valence band states. Other functional polaritonic devices are also investigated using other material elementary excitations (*8*-*11*).

The phonon is the elementary excitation of lattice vibrations in a solid; in an ionic crystal its optical mode is strongly coupled with light, leading to the formation of a phonon polariton (*12*, *13*). As vibration frequencies of many polar materials lay in the terahertz (THz) spectral region, the excitation and emission of such polaritons from a near-infrared source via a non-linear interaction has attracted much interest (*14*, *15*), including its use in THz spectroscopy applications (*15*). In addition, the phonon polariton displays very peculiar features of coherent emission and propagation (*16*, *17*), in its wave-guiding (*18*), coupling to metamaterial resonators (*19*) and tunable resonant energy scaled down to a few atomic layers in two-dimensional materials (*20*). Control and manipulation of such properties open up the possibilities to build a platform based on solid state materials to explore THz nano-photonics and -phononics.

In contrast to optical pumping of phonon polaritons, electrical pumping, desirable for numerous applications, remains a challenging task. Here we report the demonstration of an electrically pumped phonon-polariton semiconductor laser. A suitably engineered semiconductor quantum well heterostructure is electrically excited and provides gain mainly into the photon fraction of the phonon polariton. As a result, phonon polariton lasing action at an energy close to the longitudinal optical (LO) phonon energy ($\hbar\omega_{\mathrm{LO}}$) with a phonon fraction of 65% is achieved.

As shown in Fig. 1A, phonon polaritons are generated in our structure in the quantum cascade active region based on an InGaAs/AlInAs heterostructure (*21*) (the computed electron band structure is shown in Fig. 1S in the supplementary material). In this work we targeted the AlAs transverse optical phonon (TO) in the AlInAs barrier layers at $\hbar\omega_{\mathrm{TO}} \approx 42$ meV because it is energetically well separated from the frequency of the other TO phonons (InAs: $\hbar\omega_{\mathrm{TO}} \approx 29$ meV and GaAs: $\hbar\omega_{\mathrm{TO}} \approx 31$ meV) allowing it to be selectively addressed optically. The active region, based on a bound-to-continuum transition scheme (*22*), is designed to provide gain mainly into the photon fraction of the polariton at $\hbar\omega_{\mathrm{TO}}$ of AlAs by our density matrix simulator (*23*). Additionally, recent theoretical results show that, depending on frequency, the phonon fraction will also experience a gain of approximatively 1-10% of the one experienced by the photon fraction (*24*). This has been confirmed for the present design by computing the phonon polariton emission rate, shown in Fig 1C, using a non-equilibrium Green's function model (*25*). Although the inter-subband laser transition energy is 42 meV, due to the polaritonic nature of the emitted radiation the laser is expected to operate at 48.2 meV. This is lower than the peak of the emission rate since the photon loss rate is greater than the phonon one. The cavity thus favors a greater phonon component of the polariton, resulting in a non-photonic fraction of the emission rate of 7% (at the experimental lasing energy of 47.2 meV it is 10%). The simulations also show that the polariton laser threshold is lower than that expected from the purely optical emission (see the supplementary section S5).

A 4.1 μm thick structure comprised of 60 repetitions of the active region was grown by a molecular beam epitaxy on InP substrates. For comparison, we also grew six additional structures in which the gain is designed to peak at different energies higher than $\hbar\omega_{\mathrm{TO}}$ of AlAs. Waveguiding of the phonon polaritons is done in a hybrid way, as the phonon excitation is naturally confined within the AlInAs layers, while the photonic part is guided along the plane of the layers by two metallic contact layers (*26*). As shown by the computation of the mode displayed in Fig. 1A, the squared electric field is concentrated in the barrier material in a deep subwavelength dimension. The contribution of the mechanical energy $U_{\mathrm{mechanical}}$ to the total energy $U_{\mathrm{total}}$ as a function of photon energy and position in the active region is shown in Fig. 1B.

**Results**

To provide an unambiguous probe of the phonon-polaritonic nature of the emission, we investigated the latter using both a direct measurement of the emission as well as using Raman scattering with a green light excitation. First, the subthreshold emission spectra ($T = 10$ K) of a 30μm wide and 250μm long device were recorded using a home-build vacuum Fourier transform infrared spectrometer equipped with a He-cooled Si bolometer. As seen in Fig. 2A, the emission spectra display an asymmetric shape, with two sharp maxima, separated by the energy gap formed between the two AlAs LO and TO optical phonon energies, which were determined by Raman scattering (Fig. 2B) and far-infrared transmission measurements (Fig. 2C), respectively. Lasing occurs at a photon energy of 47.2 meV, which is slightly higher than $\hbar\omega_{\mathrm{LO}}$ of AlAs. The light intensity versus injected current curve (shown in Fig. 2D) exhibits a single threshold behavior. We attribute the absence of a double threshold, seen usually in exciton polariton lasers, to the different nature of the material excitations: In the case of exciton polaritons, the second threshold occurs when excitons are ionized to free electrons/holes for a large density whereas phonons are not ionized (*6*, *7*, *27*, *28*).

To probe the polariton, we measured the low temperature ($T$ = 5 K), micro Raman scattering spectra of a 300 μm long and 30 μm wide quantum cascade structure under operation. To limit the dissipation, the device was driven at a 10% duty cycle. A polarized continuous-wave green laser light with a wavelength of $\lambda$ = 532 nm was focused on the cleaved facet of the laser and the back scattered light was measured though a polarizer. As depicted in Fig. 3A, for Raman shifts below 40 meV, the spectrum was consistent with the phonon peaks associated with the InGaAs/AlInAs layers. In particular the peaks at 28 meV and 32 meV were assigned to the InAs and GaAs TO phonons, respectively. In this geometry, an AlAs TO phonon peak was not visible on the Raman spectra due to the smaller volume concentration ratio of AlAs. The peak at 35 meV denoted by "x" is an artefact attributed to the Raman scattering inside the optics used in the setup. In this spectrum, for a driving current of $I$ = 1050 mA a peak appears both in the Stokes and anti-Stokes sides at an energy corresponding exactly to the energy of the phonon-polariton. As expected, this peak disappeared when the device was driven at a current below the threshold ($I$ = 670 mA). While for the TO phonon peaks, the ratio of the Anti-Stokes to Stokes sideband intensity is 0.42 and corresponds to a thermal phonon population at a temperature of ~150 K mainly caused by localized heating from the $P$ = 20 mW green laser spot (assuming a constant Raman cross-section and thaking the InAs TO phonon). The same ratio is close to unity for the polaritonic peak. This is expected as the occupancy of the phonon polariton is well above unity above threshold. In fact, in the driving conditions considered in this experiment, we estimated a population of the order of approximately $10^5$ polaritons. As shown in Fig. 3B, the peak exhibits the selection rules expected for nonlinear optical interaction with an electric field along the growth direction. The Raman signal originating from the phonon polaritons will only be visible for the horizontal (along the plane of the layers)-horizontal polarization direction combination, while the other phonon peaks are visible for the cross-polarization combinations (vertical (perpendicular to the plane of the layers)-horizontal and horizontal-vertical).

Shown in Fig. 3C is the Raman backscattering spectra obtained for a laser where the active region consisted of a reference laser heterostructure where the AlInAs barriers were substituted by GaAsSb ones. These devices do not contain any AlInAs material, but exhibit a similar intersubband gain coefficient (*29*). Although the laser emitted at roughly the same frequency, in the Raman shift region where the phonon-polariton signal is expected, only a faint peak is visible just above the noise. This behavior is expected because the coupling of the photons to the phonon has been reduced by the much larger detuning of the photon emission to the closest TO phonon line of GaAs at 32 meV. Finally, Fig. 3D compares the intensity of the Raman peak and the light power as a function of the injected current for these two devices. In the device with the AlInAs barrier layers, the Raman peak follows linearly the intensity of the measured laser signal, in agreement with the fact that both emissions originated from the same phonon polariton. Note, however, that the Raman emission in backscattering is normally forbidden by the momentum selection rule, because of the very small momentum carried by the phonon polariton. Emission is still detected because of the very short penetration length of the pump laser. As a result, it is actually difficult to use the intensity of the Raman sideband as a true quantitative measure of the phonon weight of the polariton.

In order to further quantify the polaritonic nature of the emission, high resolution ($\Delta\nu$ = 0.015 meV) optical spectra of various Fabry Pérot (FP) lasers of fixed cavity lengths (1 mm), designed to have their optical gain peak between 53 meV and 47 meV, are reported in Fig. 4A. As evident from the data, the longitudinal mode spacing rapidly shrinks in a very small photon energy range (≈ 5 meV). The group refractive index $n_g$, experimentally retrieved from the angular frequency spacing $\Delta\omega$ of the longitudinal modes of the FP cavity using $n_g = c/v_g = \pi c / (L\Delta\omega)$ where $c$ is the light velocity and $v_g$ is the group velocity, is plotted by red circles in Fig. 4B. The group index exhibits a very strong frequency dependence, changing by a factor of two within 2 meV. Such highly dispersive $n_g$ is expected for a phonon polariton, arising from the "slowing down" of the group velocity as the upper polariton branch departs from the light line and converges to the pure phonon excitation. As a contrast, the blue triangles in Fig. 4B that report the group index $n_g$ of the reference quantum cascade laser with GaAsSb barriers show a flat dispersion. The comparison between these two characteristics shows clearly that the strong dispersion observed in the group index $n_g$ must be predominantly attributed to the presence of the AlAs optical phonon and that the contribution of the intersubband gain is negligible.

In contrast to the case of the exciton polaritons, where the dispersion $\omega(k)$ of the polaritons is measured directly by angle-dependent photoluminescence (*30*), our approach of retrieving $n_g(\omega) = c\,(\partial k/\partial\omega)$ is a measure of its inverse derivative. In Fig. 5B, we compare the experimental results to the prediction of a theoretical model (described in the section S4 in the supplementary materials), in which the light-matter coupling is studied in the dipole gauge, and the quadratic term of the vector potential appears as a polarization self-interaction (*24, 31*). In this model, the Rabi coupling energy $\Omega_R$ between the polarization arising from the phonon excitation and the fundamental cavity mode, responsible for the anti-crossing between them, can be written at resonance as

$$\Omega_R = (\omega_p/2)\, f_p^{1/2}, \quad (1)$$

where $\omega_p$ is a plasma frequency associated with the phonon excitation having a resonance energy of $\hbar\omega_o$, and $f_p$ is their filling factor of the phonon-containing material in the cavity. The polariton dispersion is obtained by solving the following equation:

$$(\omega^2 - (\omega_o^2+\omega_p^2))(\omega^2 - \omega_c^2) = f_p\omega_c^2\omega_p^2, \quad (2)$$

where $\omega_c$ the cavity photon frequency. We consider the AlAs TO phonon excitation as providing a mechanical resonance at $\omega_o = \omega_{TO}$. The effective plasma frequency of the Al and As ions is given by $\omega_p = (\omega_{LO}^2 - \omega_{TO}^2)^{1/2}$. $f_p$ is here defined as the volume fraction of AlAs as used in a dielectric function model under the effective medium approximation. The measured AlAs $\hbar\omega_{LO}$ = 45.8 meV and $\hbar\omega_{TO}$ = 43.2 meV, together with $f_p$ = 0.067, give $\Omega_R$ =2.0 meV.

Equation (2) is solved using the following approach. At an energy $\hbar\omega_A$ = 47.7 meV chosen to be much larger than the anticrossing point of the phonon and photon excitations, the slope of the dispersion is determined mostly by interband transitions (*12*). In addition, at this point, the contribution of the polarizations of InAs and GaAs is large enough to neglect that one of AlAs because of their higher volume ratios (41% for GaAs and 53% for InAs) as compared to the one of AlAs (6%). Hence the waveguide group index $n_g$ and effective index $n_{eff}$ are approximately determined by the reference QCLs in which all the AlInAs barriers are replaced by GaAsSb ones. While $n_g$ (= 4.3) is retrieved from the longitudinal mode spacing of the FP cavity laser of one of the InGaAs/GaAsSb reference structures, the effective index $n_{eff}$ (= 3.3) is derived from the grating period dependence on the emission wavelength of first order distributed feedback lasers based on the same reference structure (see in Figs. 2S in the Supplement material). Thus, the bare cavity frequency $\omega_c$ is given as a function of $k$ by:

$$\omega_c - \omega_A = (c/n_g)\,(k - \omega_A n_{eff}/c). \quad (3)$$

The computation of the polaritonic branches by solving the equations (2) and (3) enables us to compute the group index of $n_g = c\,(\partial k/\partial \omega)$ as a function of $\omega$ and compare it to the measured $n_g$ as shown by the red line in Fig. 4B. As a comparison, we also report by a dashed line in Fig. 4B the predictions of a classical model that treats the active region as a quasi-bulk material, and whose details are reported in a supplemental material section S3 (*32*). This approach completely fails to correctly predict the experimental data. The excellent agreement between the computed dispersion of $n_g$ and the measured one is further evidence of the polaritonic nature of the emission.

In Fig. 4C, the measured emission energies are plotted onto the computed upper branch of the phonon polariton dispersion. From their wave vectors, the phonon and photon fraction of the polaritons can be derived from the Hopfield coefficients $\alpha_{photon}$ and $\alpha_{phonon}$ (with $\alpha_{photon} + \alpha_{phonon} = 1$) (*31*, *33*), as shown in Fig. 4D, a phonon fraction as high as $\alpha_{phonon} = 65\%$ is inferred.

**Discussion**

The fact that the lasing is observed on the Raman scattering spectra is a strong indication of the phonon-polariton nature of the emitted radiation. Indeed, while the creation of sidebands on a near infrared carrier has been reported previously, the effect was either based on the bulk GaAs $\chi^{(2)}$ non-linearity using long interaction length and a careful phase matching (*34*) or based on a resonant near-infrared excitonic non-linearity (35). In contrast, in our case, the interaction length is very short (< 0.5 µm) and the laser is detuned. The Raman scattering is a consequence of the large phonon component of the polaritons and the large resulting optical non-linearity at the polariton energy. The large phonon component of the polariton is also responsible for the very steep dispersion of the emission which translates into a very strong energy dependence of the group index.

In our polariton laser, the threshold is reached when the overall polariton gain, sum of its photon and phonon components (*24*), equals the losses. The polariton lifetime can be written in terms of the weighted content of the photon and phonon part: $1/\tau_{polariton} = \alpha_{photon}/\tau_{cavity} + \alpha_{phonon}/\tau_{phonon}$. Considering the values for the device operating at an energy of 47.2 meV with the maximum phonon fraction of 0.65, we evaluate for the cavity a value $\tau_{cavity} = 1.7$ ps, taking into account absorption losses by the metal and the active region. The width of the TO phonon Raman peak strongly depends on the angle of the extracted light (36), and therefore does not reflect the lifetime broadening in our setup. Taking only the intrinsic broadening (36), we obtain $\tau_{polariton} = 3.2$ ps, corresponding to an energy broadening of 0.2 meV. The latter is one order of magnitude smaller than the Rabi coupling energy ($\Omega_R = 2.0$ meV), justifying the existence of the polariton description even in the absence of gain, in contrast to our previous results reported in (37). In order to lower the threshold of phonon-polariton lasing, one possibility is to use binary AlAs barriers without disorder atomic distributions as a much narrower phonon linewidth is theoretically predicted in that case (38).

In contrast to the upper polaritonic branch, we did not observe laser action in the lower polariton branch. We attribute this behavior to the larger polariton losses, arising from the proximity of the InAs and GaAs phonons.

In conclusion, we have demonstrated the first electrically pumped phonon polariton laser. Our claim is supported by the direct observation of the photon, phonon, and polariton signatures of the emission. Specifically, we have measured the photon emission into a laser mode close to the AlAs LO phonon frequency, a highly non-thermal phonon population at the laser frequency above, but not below, laser threshold, as well as a highly frequency-dependent group index, which matches very well the one predicted from a phonon-polariton dispersion and strongly deviates from that of a classical model. Finally, our simulations of the phonon-polartion emission rate show a peak at a frequency close to the observed laser frequency, and a similar threshold current density to the experimental one. A unique feature of this laser is that a large fraction of the energy of the emitted "light" is carried in the mechanical motion of the atoms, and this device can therefore be seen as a version of a phonon laser. Indeed, because of their bosonic nature, it is legitimate to consider a laser-like process in which a coherent population of phonons with an occupation number much larger than one is created and maintained by pumping. The first implementation of such an idea used as phonon modes the mechanical excitations of the $Mg^{+}$ ion in a trap potential (39). Using an opto-mechanical platform, a phonon laser operating at a phonon frequency of 21 MHz was demonstrated by optical pumping. A pure mechanical phonon laser operating at 100 kHz has also been achieved recently using a piezo-electric excitation of a micromechanical resonator (40). Much higher frequencies, achieved using optical phonons of solids, have also been considered (*41, 42*). However, as was pointed out by Chen and Khurgin (*41*), the very large optical phonon density of states makes the realization of such optical phonon lasers difficult as all the modes of the resonator within the energy range of the gain must be populated until threshold is reached for the mode with the large gain and lower losses. In the work presented here, the use of phonon polaritons alleviates this problem as the coupling of the phonon to the photon effectively decreases the density of states close to the anticrossing point, thus reducing the threshold current density. The effect is similar to the one observed in microcavities, where the coupling of the exciton to the cavity mode reduces their mass and enables their condensation. Finally, because the energy is partially stored under the form of phonons that have a relatively long lifetime and are naturally confined within nanometer thick layers, phonon polariton lasers could be operated with extremely small cavity sizes that would maintain higher quality factors than the one possible with plasmonics (43). As a result, the design concept explored here would be very well suited for applications based on two-dimensional Van der Waals materials, where low-loss Boron Nitride phonons could provide the phonon-polariton mode (44).

## References and Notes:

1. I. Carusotto, C. Ciuti, Quantum fluids of light. *Rev. Mod. Phys.* **85**, 299-366 (2013).
2. J. Kasprzak, M. Richard, S. Kundermann, A. Bass, P. Jeambrun, J. M. J. Keeling, F. M. Marchetti, M. H. Szymańska, R, André, J. L. Staehli, V. Sanova, P. B. Littlewood, B. Deveaud, Le Si Dang, Bose-Einstein condensation of exciton polaritons. *Nature* **443**, 409-414 (2006).
3. Daniele Sanvitto, Stéphane Kéna-Cohen, The road towards polaritonic devices. *Nat. Mater.* **15**, 1061-1073 (2016).
4. D. Ballarini, M. De Giorgi, E. Cancellieri, R. Houdré, E. Giacobino, R. Cingolani, A. Bramati, G. Gigli, D. Sanvitto, All-optical polariton transistor. *Nat. Commun.* **4**, 1778 (2013).

5. C. Sturm, D. Tanese, H. S. Nguyen, H. Flayac, E. Galopin, A. Lemaître, I. Sagnes, D. Solnyshkov, A. Amo, G. Malpuech, J. Bloch, All-optical phase modulation in a cavity-polariton Mach-Zehnder interferometer. *Nat. Commun.* **5**, 3278 (2014).
6. A. Imamoglu, R. J. Ram, S. Pau, Y. Yamamoto, Nonequilibrium condensates and lasers without inversion: exciton-polatiron laser. *Phys. Rev. A* **53**, 4250-4253 (1996).
7. C. Schneider, A. Rahimi–Iman, N. Y. Kim, J. Fischer, I. G. Savenko, M. Amthor, M. Lermer, A. Wolf, L. Worschech, V. D. Kulakovskii, I. A. Shelykh, M. Kamp, S. Reitzenstein, A. Forchel, Y. Yamamoto, S. Höfling, An electrically pumped polariton laser. *Nature* **497**, 348-352 (2013).
8. Raffaele Colombelli, Jean-Michel Manceau, Perspective for intersubband polariton laser. *Phys. Rev. X* **5**, 011031 (2015).
9. P. Li, X. Yang, T. W. W. Maβ, J. Hanss, M. Lewin, A. –K. Michel, M. Wuttig, T. Taubner. Reversible optical switching of highly confined phonon-polaritons with an ultrathin phase-change material. *Nat. Mater.* **15**, 870-876 (2016).
10. H. Huebl, C. W. Zollitsch, J. Lotze, F. Hocke, M. Greifenstein, A. Marx, R. Gross, S. T. B. Goennenwein, High cooperativity in coupled microwave resonator ferromagnetic insulator hybrids. *Phys. Rev. Lett.* **111**, 127003 (2013).
11. Y. Tabuchi, S. Ishino, T. Ishikawa, R. Yamazaki, K. Usami, Y. Nakamura, Hybridizing ferromagnetic magnons and microwave photons in the quantum limit. *Phys. Rev. Lett.* **113**, 083603(2014).
12. D. L. Mills, E. Burstein, Polaritons: the electromagnetic modes of media. *Rep. Prog. Phys.* **37**, 817-926 (1974).
13. C. H. Henry, J. J. Hopfield, Raman scattering by polaritons. *Phys. Rev. Lett.* **15**, 964-966 (1965).
14. J. M. Yarborough, S. S. Sussman, H. E. Purhoff, R. H. Pantell, B. C. Johnson, Efficient, tunable optical emission from $LiNbO_3$ without a resonator. *Appl. Phys. Lett.* **15**, 102-105 (1969).
15. K. Kawase, M. Sato, T. Taniuchi, H. Ito, Coherent tunable THz wave generation from $LiNbO_3$ with monolithic grating coupler. *Appl. Phys. Lett.* **68**, 2483 (1996).
16. J. J. Greffet, R. Carminati, K. Joulain, J. P. Mulet, S. Mainguy, Y. Chen, Coherent emission of light by thermal sources. *Nature* **416**, 61-64 (2002).
17. T. Feurer, J. C. Vaughan, K. A. Nelson, Spatiotemporal coherent control of lattice vibrational waves. *Science* **299**, 374-377 (2003).
18. N. S. Stoyanov, D. W. Ward, T. Feurer, K. A. Nelson, Terahertz polariton propagation in patterned materials. *Nat. Mater.* **1**, 95-98 (2002).
19. D. J. Shelton, I. Brener, J. C. Ginn, M. B. Sinclair, D. W. Peters, K. R. Coffey, G. D. Boreman, Strong coupling between nanoscale metamaterials and phonons. *Nano Lett.* **11**, 2104-2108 (2011).
20. S. Dai, Z. Fei, Q. Ma, A. S. Rodin, M. Wagner, A. S. McLeod, M. K. Liu, W. Gannett, W. Regan, K. Watanabe, T. Taniguchi, M. Thiemens, G. Dominguez, A. H. Castro Neto, A. Zettl, F. Keilmann, P. Jarillo-Herrero, M. M. Fogler, D. N. Basov, Tunable phonon polaritons in atomically thin van der Waals crystals of Boron Nitride. *Science* **343**, 1125-1129 (2014).
21. J. Faist, F. Capasso, D. L. Sivco, C. Sirtori, A. L. Hutchinson, A. Y. Cho, Quantum cascade laser. *Science* **264**, 553-556 (1994).
22. J. Faist, M. Beck, T. Aellen, E. Gini, Quantum-cascade lasers based on a bound-to-continuum transition. *Appl. Phys. Lett.* **78**, 147-149 (2001).

23. R. Terazzi, J. Faist, A density matrix model of transport and radiation in quantum cascade lasers. *New J. Phys.* **12**, 033045 (2010).
24. M. Franckié, C. Ndebeka-Bandou, K. Ohtani, and J. Faist, *Phys. Rev. B* **97**, 075402 (2018).
25. A. Wacker, M. Lindskog, D.O. Winge, Nonequilibrium Green's function model for simulation of quantum cascade laser devices under operating conditions. *IEEE J. Sel. Top. Quant. Electron.* **19** 1200611-1 – 1200611-11 (2013)
26. K. Unterrainer, R. Colombelli, C. Gmachl, F. Capasso, H. Y. Hwang, A. Sergent, A. Michael, D. L. Sivco, A. Y.Cho, Quantum cascade lasers with double metal-semiconductor waveguide resonators. *Appl. Phys. Lett.* **80**, 3060-3062 (2002).
27. H. Ding, G. Weihs, C. Santori, J. Bloch, Y. Yamamoto, Condensation of semiconductor microcavity exciton polaritons. *Science* **298**, 199-202 (2002).
28. H. Ding, G. Weihs, D. Snoke, J. Bloch, Y. Yamamoto, Polariton lasing vs. photon lasing in a semiconductor microcavity. *Proc. Natl. Acad. Sci. USA* **100**, 15318-15323 (2003).
29. K. Ohtani, M. Beck, M. J. Süess, J. Faist, A. M. Andrews, T. Zederbauer, H. Detz, W. Schrenk, G. Strasser, Far-infrared quantum cascade lasers operating in AlAs phonon reststrahlen band. *ACS Photonics* **3**, 2280-2284 (2016).
30. R. Houdré, C. Weisbuch, R. P. Stanley, U. Oesterle, P. Pellandini, M. Ilegems, Measurement of cavity-polariton dispersion curve from angle-resolved photoluminescence experiments. *Phys. Rev. Lett.* **73**, 2043-2046 (1994).
31. Y. Todorov, C. Sirtori, Intersubband polaritons in the electrical dipole gauge. *Phys. Rev. B* **85**, 045304 (2012).
32. H. Chu, Y. -C. Chang, Phonon-polariton modes in superlattices: The effect of spatial dispersion. *Phys. Rev. B* **38**, 12369-12376 (1988).
33. J. J. Hopfield, Theory of the contribution of excitons to the complex dielectric constant of crystals. *Phys. Rev.* **112**, 1555-1567 (1958).
34. S. S. Dhillon, C. Sirtori, J. Alton, S. Barbieri, A. De Rossi, H. E. Beere, and D. A. Ritchie, Terahertz transfer onto a telecom optical carrier. *Nat. Photonics* **1**, 411–415, (2007).
35. S. Houver, P. Cavalié, M. R. St-Jean, M. I. Amanti, C. Sirtori, L. H. Li, A. G. Davies, E. H. Linfield, T. A. S. Pereira, A. Lebreton, J. Tignon, and S. S. Dhillon, Optical sideband generation up to room temperature with mid-infrared quantum cascade lasers. *Opt. Express*, **23**, 4012–4019 (2015).
36. M. Giehler, E. Jahna, K. Ploog, Linewidth of phonon modes in infrared transmission spectra of GaAs/AlAs superlattices. *J. Appl. Phys.* **77**, 3566-3568 (1995)
37. K. Ohtani, C. Ndebeka-Bandou, L. Bosco, M. Beck, J. Faist, arXiv:1610.00963.
38. A. Debernardi, Phonon linewidth in III-V semiconductors from density-functional perturbation theory. *Phys. Rev. B* **57**, 12847-12858 (1998).
39. K. Vahala, M. Herrmann, S. Knünz, V. Batteiger, G. Saathoff, T. W. Hänsch, Th. Udem, A phonon laser. *Nat. Phys.* **5**, 682-686 (2009).
40. I. Mahboob, K. Nishiguchi, A. Fujiwara, H. Yamaguchi, Phonon lasing in an electromechanical resonator. *Phys. Rev. Lett.* **110**, 127202 (2013).
41. J. Chen, J. B. Khurgin, Feasibility analysis of phonon lasers. *IEEE J. Quantum Electron.* **39**, 600-607 (2003).
42. J. B. Khurgin, S. Bajaj, S. Rajan, Amplified spontaneous emission of phonons as a likely mechanism for density-dependent velocity saturation in GaN transistor. *Appl. Phys. Exp.* **9**, 094101 (2016).

43. J. D. Caldwell, L. Lindsay, V. Giannini, I. Vurgaftman, T. L. Reinecke, S. A. Maier, O. J. Glembocki. Low-loss, infrared and terahertz nanophotonics using surface phonon polaritons. *Nanophotonics* **4**, 1-26 (2014).
44. A. J. Giles, S. Dai, I. Vurgaftman, T. Hoffman, S. Liu, L. Lindsay, C. T. Ellis, N. Assefa, I. Chatzakis, T. L. Reinecke, J. G. Tischler, M. M. Fogler, J. H. Edgar, D. N. Basov, and J. D. Caldwell, Ultra-low loss polaritons in isotopically pure boron nitride. *Nat. Mat.* **17**, 134-139 (2018).
45. C. Deutsch, M. Krall, M. Brandstetter, H. Detz, A. M. Andrews, P. Klang, W. Schrenk, G. Strasser, K. Unterrainer. High performance InGaAs/GaAsSb terahertz quantum cascade lasers operating up to 142 K. *Appl. Phys. Lett.* **101**, 211117 (2012).
46. G. Scamarcio, L. Tapfer, W. König, A. Fischer, K. Ploog, E. Molinari, S. Baroni, P. Giannozzi, and D. de Gironcoli, Infrared reflectivity by transverse-optical phonons in $(GaAs)_m/(AlAs)_n$ ultrathin-layer superlattices. *Phys. Rev. B* **43**, 14754-14757 (1991).
47. O. Madelung, U. Rössler, and M. Schulz, eds., "Indium arsenide (InAs) dielectric constants" in *Group IV elements, IV-IV and III-V compounds. Part a – lattice properties* (Springer Berlin Heidelberg, Berlin, Heidelberg, 2001) pp. 1-3.
48. D. J. Lockwood, G. Yu, and N. L. Rowell, Optical phonon frequencies and damping in AlAs, GaP, GaAs, InP, InAs and InSb studied by oblique incidence infrared spectroscopy. *Sol. Stat. Comm.* **136**, 404-409 (2005).
49. O. Madelung, U. Rössler, and M. Schulz, eds., "Aluminum arsenide (AlAs) dielectric constants" in *Group IV elements, IV-IV and III-V compounds. Part a - lattice properties* (Springer Berlin Heidelberg, Berlin, Heidelberg, 2001) pp. 1-4.
50. M. A. Ordal, R. J. Bell, R. W. Alexander, Jr., L. L. Long, and M. R. Querry, Optical properties of fourteen metals in the infrared and far infrared: Al, Co, Cu, Au, Fe, Pb, Mo, Ni, Pd, Pt, Ag, Ti, V and W, *Appl. Opt.* **24,** 4493 (1985).

**Acknowledgments:**

We acknowledge very fruitful discussions with Angela Vasanelli, Simone de Liberato, and Jacob B. Khurgin, and also thank Aaron Maxwell Andrews, Tobias Zederbauer, Hermann Detz, Werner Schrenk, and Gottfried Strasser for supporting MBE growth of InGaAs/GaAsSb material at early stage of this work**. Funding:** The presented work is supported in part by the ERC Advanced grant Quantum Metamaterials in the Ultra Strong Coupling Regime (MUSiC) with the ERC Grant 340975, as well as by the Swiss National Science Foundation (SNF) through the National Center of Competence in Reasearch Quantum Science and Technology (NCCR QSIT). **Author contributions:** KO did the design of all structures, the MBE growth of Sb-based material, the computation of the theoretical polariton dispersion and dielectric function calculations, the device characteristics, and the low temperature Raman scattering experiment under biasing. BM has conducted recent Raman experiments and LB processed the grown structures into laser devices. MF, CNB, and JF have developed the theoretical framework and software for the phonon-polariton gain calculations. MB performed MBE growth of the Sb-free structures. KO and JF had the original idea and have drafted the initial manuscript, and MF and BM have done the reviewing and editing. Finally, JF has been supervising this work. **Competing interests:** Authors declare no competing interests. **Data and materials availability:** All data is available in the main text or the supplementary materials. The measured samples and the codes used to generate the theoretical data are available at ETH Zürich.

**Supplementary Materials:**

Materials and Methods

Figures S1-S3

Table S1

References (*45-50*)

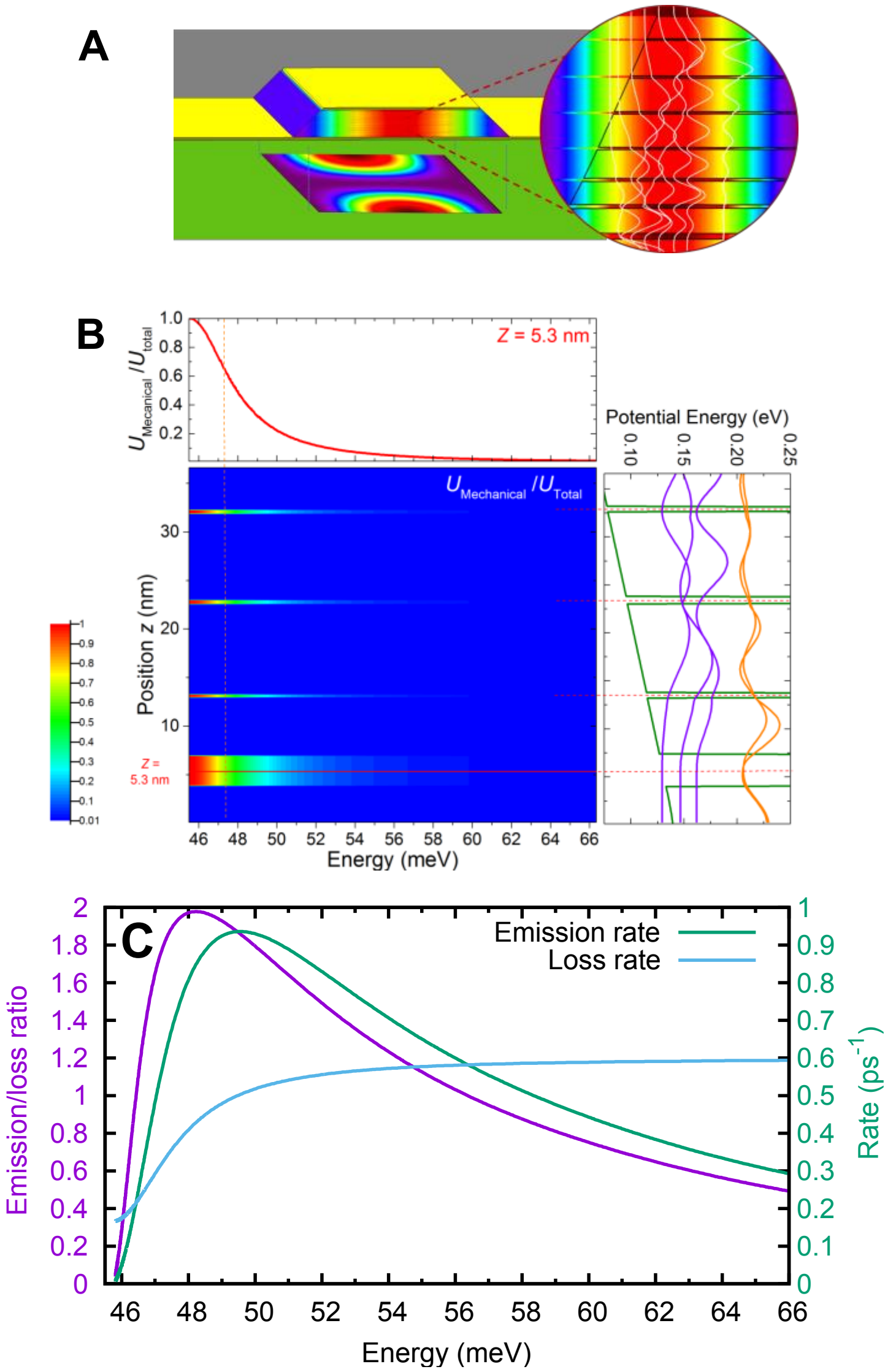

**Fig. 1**. **Schematics and design of the device structure.** (**A**) Schematic of the phonon-polariton laser. In the waveguide the Au metal layers sandwich the quantum cascade active region and guide the photon part of the phonon polariton with a confinement factor of ≈ 1. In the magnified view, a part of the computed heterostructure potential with the relevant electron wavefunctions in the active region (depicted by white lines) is overlaid on the computed squared electric field of the fundamental cavity mode. Being associated with the AlAs vibrational mode, the phonon part of the phonon polariton is confined in the AlInAs barriers. As shown by the displayed mode, the squared electric field is concentrated in the barrier material in a deep subwavelength dimension. (**B**) The contribution of the mechanical energy $U_{\text{mechanical}}$ to the total one $U_{\text{total}}$ as a function of photon energy and position in the active region. The upper figure shows the ratio of $U_{\text{mechanical}}/U_{\text{total}}$ at $z$ = 5.3 nm. The dotted vertical line indicates the observed lasing frequency of the device. The right figure shows the corresponding position in the active region with the relevant wavefunctions overlaid. (**C**) The polariton emission and loss rates (left axis), as well as the ration between the emission and loss rates (right axis) computed at an electric field of 19 kV/cm and current density of 10.4 kA/cm$^2$. The gain is computed by first solving the transport in a non-equilibrium Green's function model and using these results to obtain the polariton coupling constants (see section S5 of the supplementary materials). Since the calculated photon lifetime $\tau_{\text{phot.}}$ = 1.7 ps is shorter than the phonon lifetime $\tau_{\text{phon.}}$ = 6 ps, the laser will operate at a lower frequency (48.2 meV) than that of peak polartion emission (49.6 meV), where the phonon component of the polariton is larger.

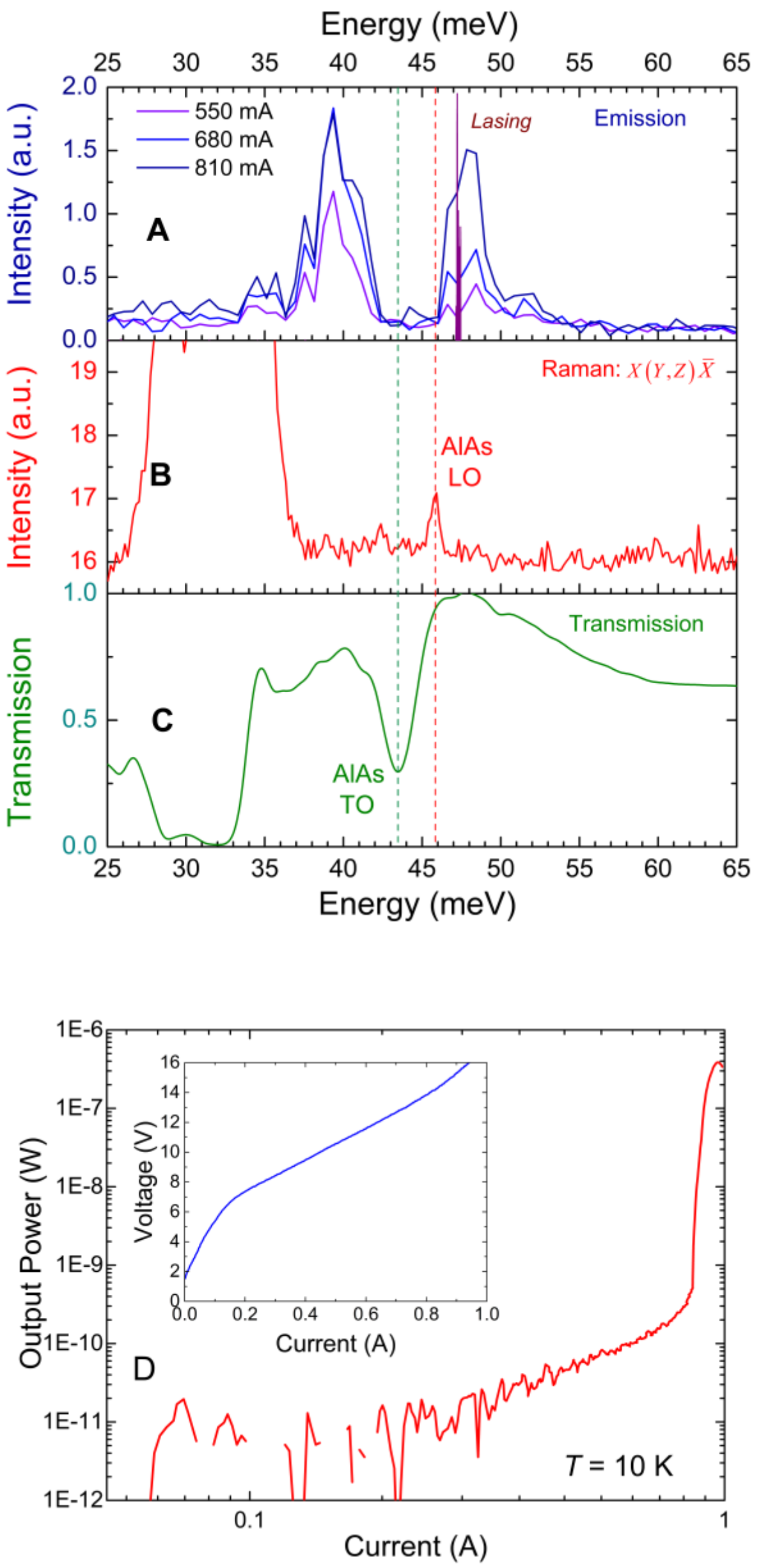

**Fig. 2. Device characteristics.** (**A**) Measured subthreshold emission spectra ($T$ = 10 K). Two peaks separated by the AlAs reststrahlen band were observed. Those peaks stem from emission from the lower and upper phonon polariton branches. (**B**) Measured Raman spectrum taken from the active region in the $X(Y,Z)\bar{X}$ geometry. (**C**) Room temperature direct transmission spectrum using a room temperature Deuterated Tri Glycine Sulfate (DTGS) detector. (**D**) Light output versus current characteristic. A single threshold current (= 880 mA) was observed. The inset shows the transport curve of the device. The ridge width of the device was 30 μm and the length was 250 μm.

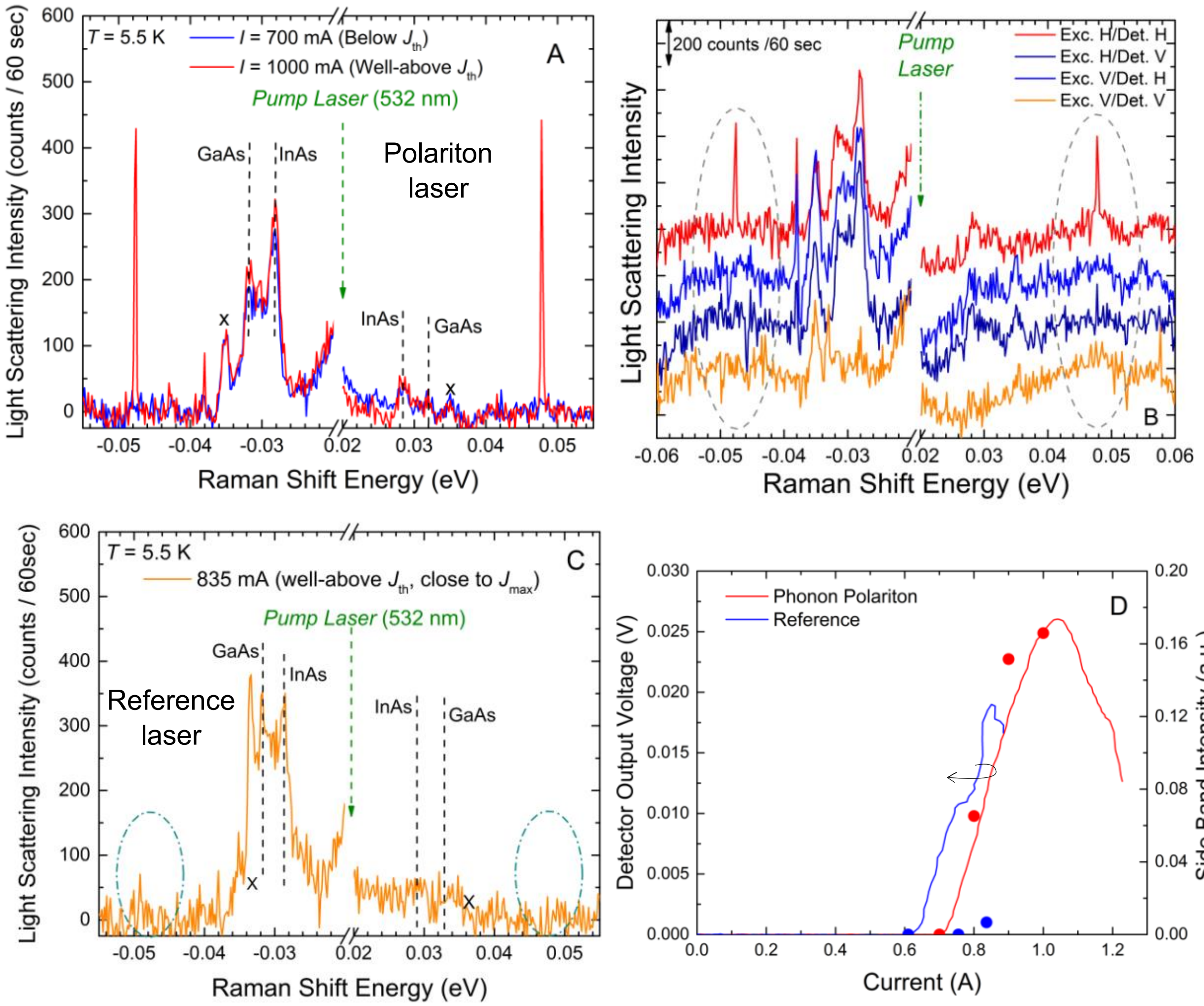

**Fig. 3** (**A**) Low temperature ($T$ = 5.5 K) polarized Raman scattering spectra of the facet of the phonon-polariton laser under operation. The peaks at 28 meV and 32 meV are attributed to the InAs and GaAs phonons of the active region, respectively. The phonon-polariton peak at 48 meV is seen when the device is driven above threshold (1060 mA) and disappears when the device is below threshold (670 mA). (B) Polarization selection rules for the phonon-polariton peak, which appears only for the horizontal-horizontal polarization combination. (C) Raman spectrum of the reference sample at a current of 835 mA, which is close to the roll-over current of the device. (D) Light and Raman side band intensity of the phonon-polariton and the reference devices as function of the injected current.

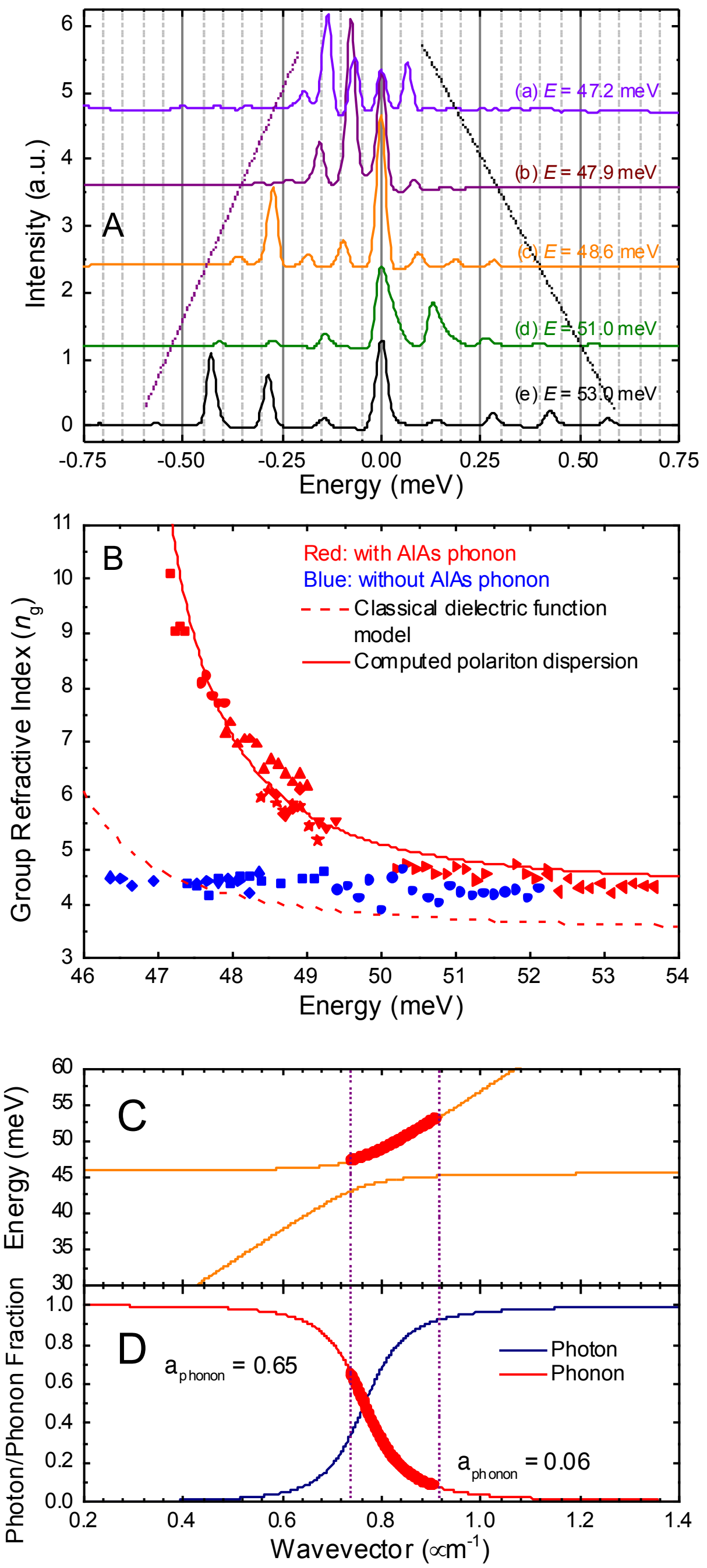

Figure 4

**Fig. 4. Phonon polariton laser emission spectra.** (**A**) High resolution emission spectra of the Fabry-Pérot (FP) devices. The cavity length of all the devices was fixed at 1.0 mm. The longitudinal mode spacing is strongly reduced when the laser emission energy approaches to the AlAs optical phonon energy. The spectral measurements were done at $T$ = 15 K. (**B**) Group

refractive index ($n_g$) retrieved from the longitudinal mode spacing of the FP devices. The red solid points represent $n_g$ of the FP devices containing the AlAs phonon oscillators while the group index $n_g$ of the FP devices without the AlAs phonon oscillators are depicted by the blue solid points. The red solid line shows $n_g$ derived from the computed phonon polariton dispersion. (**C**) Computed dispersion of the two polariton branches, including the contribution from the two phonon modes (AlAs, InAs and GaAs). The point A is the energy at which the values of the effective and group indices are experimentally measured. (**D**) Computed phonon polariton dispersion for our laser devices. The thick red lines represent the range of emission energies of the measured devices. (**C**) Computed Hopfield coefficients. In this model, a phonon fraction of 0.65 was obtained at the lowest emission energy (47.2 meV).

# Supplementary Materials for

## An electrically pumped phonon polariton laser

Keita Ohtani, Bo Meng, Martin Franckié, Lorenzo Bosco, Camille Ndebeka-Bandou,
Mattias Beck, Jérôme Faist

**This PDF file includes:**

Materials and Methods
Supplementary Text
Figs. S1 to S3
Table S1

## Materials and Methods

Materials

The samples were grown by molecular beam epitaxy on Fe-doped InP (001) substrates, starting with a 60 nm thick $In_{0.53}Ga_{0.47}As$ buffer layer. Then a 4.1 μm thick active region was deposited and the growth was completed by a 15 nm thick Si-doped *n*-type $Ga_{0.53}In_{0.47}As$ contact layer ($n = 5.0 \times 10^{18}$ cm$^{-3}$). In the active region, the following layer structure (EV2128) was repeated by 60 times: in nanometers, **3.0**/6.0/**0.3**/9.3/**0.4**/8.9/**0.4**/8.8/**0.4**/7.7/**0.5**/6.3/**0.8**/6.6/**1.4**/7.5, where $Al_{0.48}In_{0.52}As$ barrier layers are in bold, $In_{0.53}Ga_{0.47}As$ well layers are in roman, and Si-doped $In_{0.53}Ga_{0.47}As$ layer ($n = 4.1 \times 10^{17}$ cm$^{-3}$) is underlined. All the layer structures presented here are summarized in Table S1. The corresponding conduction band diagram of the one period of the active region with moduli-squared relevant wave functions is also shown in Fig. S1.

Methods

For laser device fabrication, a bottom metal contact layer composed of a 10 nm thick Titanium (Ti) and a 500 nm thick gold (Au) was first deposited on the grown wafer. It was wafer-bonded onto the Au evaporated *n*-InP (100) carrier substrate by thermo-compression bonding technique. Then, the Fe-doped InP substrate used for the epitaxial growth was removed by mechanical polishing and selective wet chemical etching. After evaporation of a 260 nm thick Ti/Au top contact, the ridge laser structures were defined by a wet chemical etching. After device fabrication, the processed wafer was cleaved into 0.25-1.0 mm long Fabry-Pérot lasers with a ridge width of 30 μm, and then mounted on copper mounts. The laser devices were placed in a temperature controlled He flow cryostat with KRS5 vacuum windows.

For the low temperature ($T$ = 5 K) micro-Raman scattering experiment, a continuous wave frequency doubled Nd:YAG (wavelength: 532.1 nm) laser was focused through the cryostat window into a 2.5 μm diameter spot by a ×50 objective lens with a numerical aperture of 0.4. Backward scattered light was collected and sent to the single grating monochrometer equipped with a $LN_2$ cooled Si CCD detector. According to the Raman scattering selection rule for a zinc blende crystal, $X(Y,Z)\bar{X}$ and $Y'(X,Z')\bar{Y}'$ geometries were used for LO and TO phonons, respectively. Here, $X$ denotes the $[100]$ axis, which is parallel to the InP substrate surface (100), and $Y$ ($Z$) is along the direction of $[010]$ ($[001]$). Also, $Y'$ and $Z'$ denote the $[011]$ and the $[0\bar{1}1]$ axes, repsectively. The power of the laser was varied from 5 mW to a maximum of 20 mW, limited by local heating at the focal point.

To determine the AlAs TO phonon energy, a normal incident light transmission was measured. The active region film was glued on a Si substrate. To compensate the temperature difference of the two experiments, the transmission spectrum was shifted by 0.8 meV, which is consistent with the temperature-induced energy shift of the LO phonon energy as measured by Raman scattering experiment.

The sub-threshold electroluminescence measurements ($T$ = 10 K) were performed using a home-made vacuum Fourier Transform Infrared (FT-IR) spectrometer equipped with a calibrated Si bolometer. A current pulse train composed of 100 ns wide pulses with a repetition rate of 1 MHz (a duty cycle of 10%) were modulated at 450 Hz to match the frequency response of the calibrated Si bolometer. The light output power was also measured by the Si bolometer. For high-resolution spectral measurements, a Bruker vacuum FT-IR was used. To minimize the broadening of the spectra, the length of the electrical pulses was kept below 50 ns. The spectral resolution used in this study was 0.125 $cm^{-1}$.

## Supplementary Text

S1. Layer structures and emission characteristics

In Table S1, the grown layer structures including thickness and chemical composition are summarized. The QCL structures are based on the $In_{0.53}Ga_{0.47}As/Al_{0.48}In_{0.52}As$ material while the reference ones are composed of the AlAs-free $In_{0.53}Ga_{0.47}As/GaAs_{0.51}Sb_{0.49}$ material. Here, $Al_{0.48}In_{0.52}As$ barrier layers are in bold, $In_{0.53}Ga_{0.47}As$ well layers are in roman, $GaAs_{0.51}Sb_{0.49}$ barrier lays are in italic, and Si-doped $In_{0.53}Ga_{0.47}As$ layer ($n = 4.1 \times 10^{17}$ $cm^{-3}$) is underlined. All the samples used here were grown on Fe-doped (001) InP substrates by molecular beam epitaxy. The computed conduction band diagrams together with the relevant electron wavefunctions of (A) EV2128 and (B) EP1561 are depicted in Fig. S1. Here, an electric field of 19.5 kV/cm is applied to align subbands. The band calculation was done by self-consistent Poisson and Schrödinger solver. The well layers are composed of lattice matched $In_{0.53}Ga_{0.47}As$ to InP, while the barrier layers are lattice matched $Al_{0.48}In_{0.52}As$ (A) or $GaAs_{0.51}Sb_{0.49}$ layers (B). Effective masses of 0.043 $m_0$ for $In_{0.53}Ga_{0.47}As$, 0.076 $m_0$ for $Al_{0.48}In_{0.52}As$, and 0.045 $m_0$ for $GaAs_{0.51}Sb_{0.49}$ ($m_0$ is a free electron mass) are used. The conduction band offset energy is 0.52 eV for (A) $Al_{0.48}In_{0.52}As$ and 0.36 eV for (B) $GaAs_{0.51}Sb_{0.49}$ (*45*). Other structures have similar band structures. Also low temperature threshold current densities and emission wavelength are summarized in Table S1.

S2. Laser characteristics of the first order distributed feedback laser based on InGaAs/GaAsSb

First order distributed feedback (DFB) laser structures equipped with a lateral grating and a quarter wave shift, as depicted in Fig. S2 A, were used for the measurements of the effective refractive index $n_{eff}$. The wafer used was EP1562 listed in Table S1. The DFBs show single mode emission with a threshold current density of ≈ 350 mA (see Figs. S2 B and C). As seen in Fig. S2 D, $n_{eff} = 3.3$ was obtained from a relationship between the lasing wavelengths and the grating periods, which is in good agreement with the simulated results by the commercial software (COMSOL Multiphysics 5.2a). A group index of $n_g = 4.3$ is measured by the laser mode spacing of the Fabry-Pérot laser as depicted in Fig. 2S E.

S3. Computed results by a classical dielectric function

We consider the structure as a uniaxial crystal with an average dielectric constant and write the dielectric constant ($\varepsilon_z$) for a direction ($z$) perpendicular to the layers in the framework of the effective media approximation (*32*, *46*):

$$\frac{1}{\langle \varepsilon_z \rangle} = \left( \frac{1}{d_{total}} \right) \left( \frac{d_{well}^{total}}{\varepsilon_{well}} + \frac{d_{barrier}^{total}}{\varepsilon_{barrier}} \right). \quad (1)$$

where $d_{well}^{total}$ $(d_{barrier}^{total})$ is the total thickness of the $In_{0.53}Ga_{0.47}As$ well ($Al_{0.48}In_{0.52}As$ barrier) layers, $\varepsilon_{well}$ ($\varepsilon_{barrier}$) is the dielectric constant of the $In_{0.53}Ga_{0.47}As$ well ($Al_{0.48}In_{0.52}As$ barrier) layer, $d_{total}$ is the total thickness of wells and barriers $(d_{total} = d_{well}^{total} + d_{barrier}^{total})$. $\varepsilon_{well}$ and $\varepsilon_{barrier}$ of the alloy materials are given by the following formulae:

$$\varepsilon_{well}\left(\omega\right)=\varepsilon_{InGaAs}^{\infty}+x_{In}\varepsilon_{InAs}^{\infty}\frac{\left(\omega_{LO(InAs)}^{2}-\omega_{TO(InAs)}^{2}\right)}{\left(\omega_{TO(InAs)}^{2}-\omega^{2}\right)}+x_{Ga}\varepsilon_{GaAs}^{\infty}\frac{\left(\omega_{LO(GaAs)}^{2}-\omega_{TO(GaAs)}^{2}\right)}{\left(\omega_{TO(GaAs)}^{2}-\omega^{2}\right)}\quad\left(x_{In}+x_{Ga}=1\right),\quad(2)$$

$$\varepsilon_{barrier}\left(\omega\right)=\varepsilon_{AlInAs}^{\infty}+x_{In'}\varepsilon_{InAs}^{\infty}\frac{\left(\omega_{LO(InAs)}^{2}-\omega_{TO(InAs)}^{2}\right)}{\left(\omega_{TO(InAs)}^{2}-\omega^{2}\right)}+x_{Al}\varepsilon_{AlAs}^{\infty}\frac{\left(\omega_{LO(AlAs)}^{2}-\omega_{TO(AlAs)}^{2}\right)}{\left(\omega_{TO(AlAs)}^{2}-\omega^{2}\right)}\quad\left(x_{In'}+x_{Al}=1\right),\quad(3)$$

where $x_i$ expresses the group III atom compositional ratio (in this case, $x_{In}$ = 0.53, $x_{Ga}$ = 0.47, $x_{In'}$ = 0.52, $x_{Al}$ = 0.48), $\varepsilon_{InGaAs}^{\infty}$ ($\varepsilon_{AlInAs}^{\infty}$) is the high frequency dielectric constant of $In_{0.53}Ga_{0.47}As$ ($Al_{0.48}In_{0.52}As$) defined as $\varepsilon_{InGaAs}^{\infty}=x_{In}\varepsilon_{InAs}^{\infty}+x_{Ga}\varepsilon_{GaAs}^{\infty}\left(\varepsilon_{AlInAs}^{\infty}=x_{In'}\varepsilon_{InAs}^{\infty}+x_{Al}\varepsilon_{AlAs}^{\infty}\right)$. The longitudinal and transverse optical phonon energies of InAs, GaAs, and AlAs of the QCL active region were determined by low temperature ($T$ = 5 K) polarized Raman scattering measurements. Because those energies did not show structure dependence, the following energies were used: $\hbar\omega_{TO(InAs)}$ = 27.5 meV, $\hbar\omega_{LO(InAs)}$ = 28.4 meV, $\hbar\omega_{TO(GaAs)}$ = 31.3 meV, $\hbar\omega_{LO(GaAs)}$ = 33.2 meV, $\hbar\omega_{TO(AlAs)}$ = 43.2 meV, and $\hbar\omega_{LO(AlAs)}$ = 45.8 meV. For high-frequency dielectric constants, the literature values of $\varepsilon_{InAs}^{\infty}=12.3$ (*47*), $\varepsilon_{GaAs}^{\infty}=10.9$ (*48*), and $\varepsilon_{AlAs}^{\infty}=8.2$ (*49*) were used. A dispersion of the Au dielectric function is also taken into account (*50*). In Fig. S3, the red dotted straight line represents the computed $n_g$ using $\varepsilon_{GaAs,AlAs,InAs}^{\infty}$ of the literature. The computed result based on the classical dielectric model does not reproduce the steep dispersion of $n_g$ experimentally observed using the experimental parameters.

## S4. Phonon-photon interaction Hamiltonian in the dipole gauge

Here we briefly describe how to derive the phonon-photon coupling energy from the theory developed for an intersubband polariton in a dipole gauge by Y. Todorov and C. Sirtori (*31*). Details can be found in Ref. (*24*). The interaction Hamiltonian $\hat{H}_{\text{int}}$ in the electrical dipole gauge is written, neglecting the magnetic interaction,

$$\hat{H}_{\text{int}}=\int\frac{d^{3}\boldsymbol{r}}{\varepsilon_{0}\varepsilon\left(z\right)}\left[-\hat{D}\left(\boldsymbol{r}\right)\cdot\hat{P}\left(\boldsymbol{r}\right)+\frac{1}{2}\hat{P}^{2}\left(\boldsymbol{r}\right)\right],\quad(1)$$

where $\hat{D}(\boldsymbol{r})$ is the displacement field, $\hat{P}(\boldsymbol{r})$ is the polarization density operator of the phonon, and $\varepsilon(z)$ is the background dielectric function in the $z$-direction (crystal growth direction). Due to a selection rule of the intersubband transitions, the polarization of the photonic part is Transverse Magnetic (TM). The two metal plates, which are separated by the active region (of thickness $L_{cav}$) and have an area $S$, guide the fundamental mode ($TM_0$) with a field confinement factor of ≈ 1. The mode profile function is defined as $f_p(z)$, and in the case of perfect metallic boundaries, the $TM_0$ mode satisfies $f_p(z)=1$, $\int_{-\infty}^{\infty}f_p(z)dz=L_{cav}$. The displacement field operator is thus expressed as

$$\hat{D}(z,r)=i\sum_{q}f_{p}(z)\sqrt{\frac{\varepsilon\varepsilon_{0}\hbar\omega_{cq}}{2SL_{cav}}}e^{iq\cdot r}\left(a_{q}-a_{-q}^{\dagger}\right),\quad(2)$$

where $\omega_{cq}$ is an angular frequency of the guided mode with an in-plane wavevector of $q$, and $a_q^{\dagger},a_q$ are photon creation and annihilation operators. The polarization is given by

$$\hat{P}(z,r) = \frac{\hbar e^*}{2SM} \sum_q \frac{\xi_\alpha(z)}{\omega_{TO}} e^{iq\cdot r} \left( d^\dagger_{\alpha-q} + d_{\alpha q} \right). \tag{6}$$

where $d_{\alpha q}$ is the TO phonon annihilation operator, $e^*$ is the effective charge of the ions, $M$ is their reduced mass, and the phonon microcurrent is defined by

$$\xi_\alpha(z) = \psi_\lambda \frac{d\psi_\mu}{dz} - \psi_\mu \frac{d\psi_\lambda}{dz} = \sqrt{\frac{2}{\pi}} \frac{1}{\sigma^2} e^{-\left(\frac{z}{\sigma}\right)^2}. \tag{5}$$

via the ground and first excited states of the harmonic oscillator potential:

$$\psi_\lambda = \left( \frac{1}{\sigma^2 \pi} \right)^{1/4} e^{\frac{-z^2}{2\sigma^2}},\ \psi_\mu = \left( \frac{1}{\sigma^2 \pi} \right)^{1/4} \sqrt{\frac{2}{\sigma^2}} z e^{\frac{-z^2}{2\sigma^2}}. \tag{3}$$

Here, $\sigma$ measures the extent of the polarization due to the presence of the phonon modes inside a material layer. We relate this quantity to the thickness $d$ of the layer by equating the filling factor $f_\alpha^0$ (Eq. (27) of Ref. (*24*)) with $d/L_{cav}$, which results in $\sigma = d/\sqrt{(2\pi)}$. After a Bogoliubov transformation for diagonalizing the quadratic Hamiltonian, we finally obtain

$$\int \frac{1}{\varepsilon\varepsilon_0} \hat{D} \cdot \hat{P} dr^3 = \sum_q \frac{\hbar\omega_P}{2} \sqrt{f^0_\alpha \frac{\omega_{cq}}{\omega_{LO}}} \left( a_q - a^\dagger_{-q} \right) \left( p^\dagger_{\alpha-q} + p_{\alpha q} \right) = \sum_q \hbar\Omega \left( a_q - a^\dagger_{-q} \right) \left( p^\dagger_{\alpha-q} + p_{\alpha q} \right), \tag{12}$$

where $p_{\alpha q}$ is the LO phonon annihilation operator and

$$\Omega = \frac{\omega_P}{2} \sqrt{\frac{\omega_{cq}}{\omega_{LO}} f^o_\alpha} \tag{13}$$

is the light-matter coupling constant, which is similar to the expression for $\Omega$ of the intersubband polariton in the dipole gauge (*31*). Here, $\omega_p^2 = \omega_{LO}^2 - \omega_{TO}^2$ is the phonon plasma frequency and is deduced from the measured phonon frequencies. Hence, the Rabi coupling energy (Eq. (1) in the main text), expressing the minimum splitting of the two blanches, is given by

$$\Omega_R = \frac{\omega_P}{2} \sqrt{f^o_\alpha}. \tag{14}$$

S5. Transport and gain simulations

The structure EV2128 has been simulated using a non-equilibrium Green's function (NEGF) model (*25*), and the resulting current-bias is shown in Fig. S3 A. There is a good agreement with the experimental data, considering the simulated optical gain is starting to surpass the calculated losses for a current density of 10.4 kA/cm$^2$, which is close to the observed threshold current density of 10.3 kA/cm$^2$ (which in addition includes a phononic contribution to the gain). The gain spectra are also consistent with the EL spectra in Fig. 2 in the main text considering the large phonon optical absorption between 42-46 meV photon energy, and which also show a blue shift with increasing injected current. We then take the results from the NEGF transport simulation to calculate the phonon-polariton emission rate as in Ref. (*24*), resulting in the emission rate shown in Fig. 1 C in the main text.

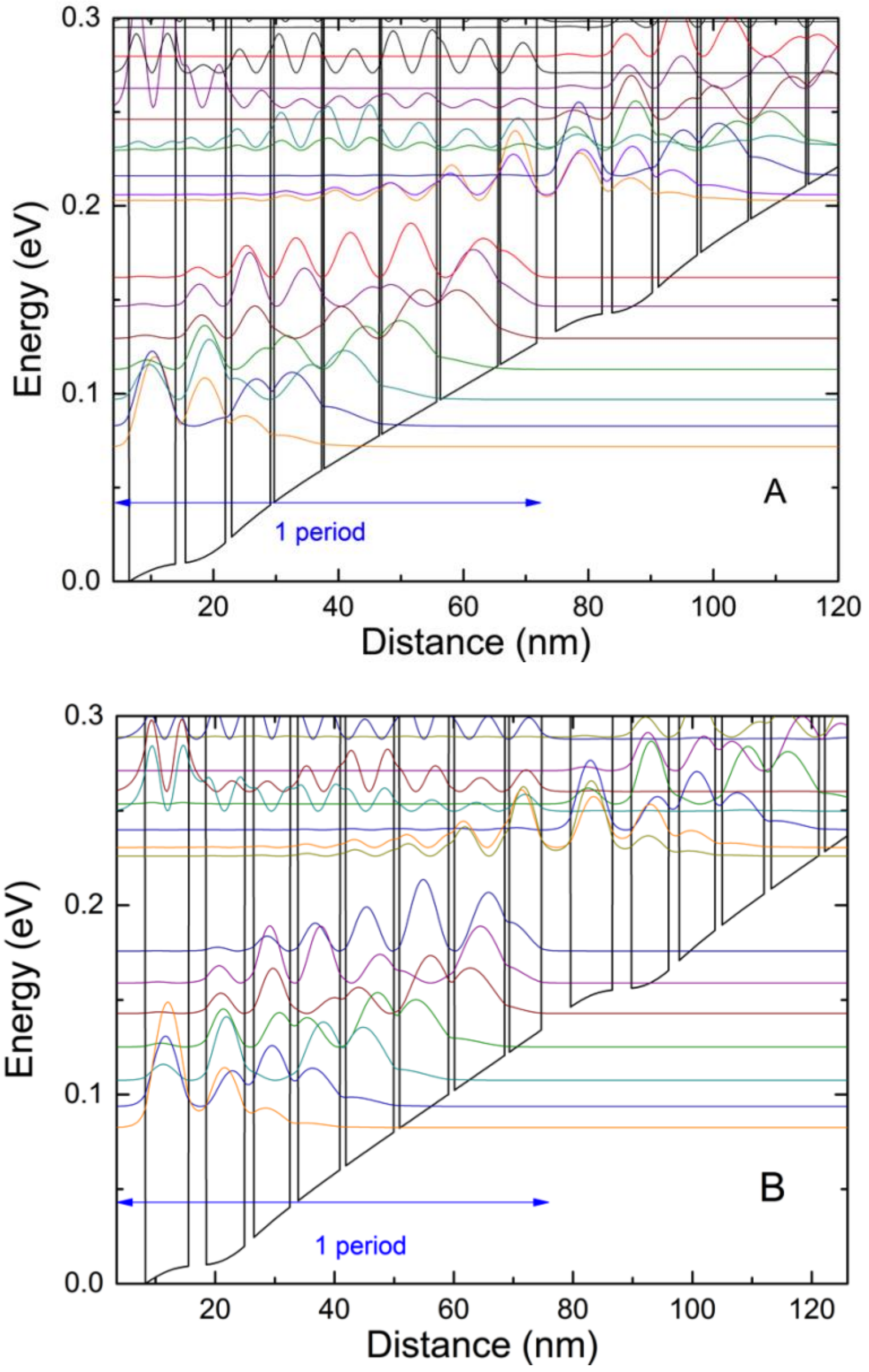

**Fig. S1 Computed conduction band diagram with relevant electron wavefunctions of a part of the active region (A) EV2128 and (B) EP1561.** An electric field of 19.5 kV/cm is applied to align the subbands.

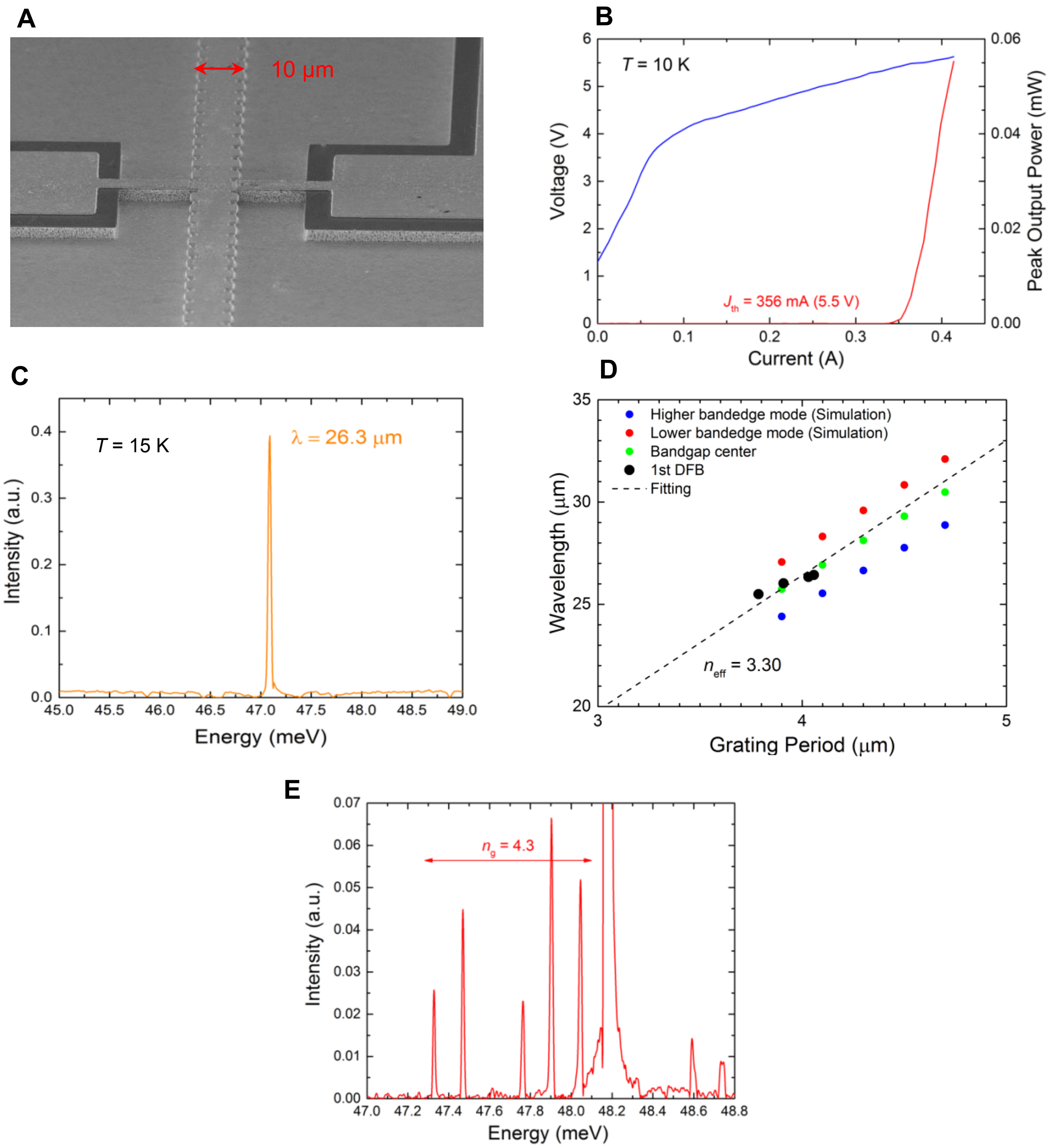

**Fig. S2 Characteristics of the first order distributed feedback (DFB) lasers and the Fabry-Perot (FP) lasers of the reference InGaAs/GaAsSb structure.** (**A**) SEM image of the fabricated first order DFB. (**B**) Voltage-current-light output characteristic of the 1st order DFB laser ($T$ = 10 K). (**C**) Single mode laser emission of the 1st order DFB laser ($T$ = 15 K) (**D**) Emission wavelength versus grating period of the 1st order DFB laser. $n_{eff}$ = 3.3 is derived. The red, blue, and green solid circles are the simulated results by COMSOL Multiphysics 5.2a. (**E**) Laser emission spectrum of the FP laser. $n_g$ of 4.3 around an energy of 47.7 meV is determined.

- $In_{0.53}Ga_{0.47}As/Al_{0.48}In_{0.52}As$ QCL layer structures

| ID | Layer structure in nanometer | $J_{th}$ & λ<br>(Low Temp) |
|---|---|---|
| EV2036 | **3.0**/5.4/**0.4**/8.6/**0.5**/8.2/**0.5**/8.1/**0.6**/7.1/**0.7**/6.1/**1.1**/6.4/**2.0**/7.2 | 6.2 kA/cm$^2$<br>23.6 µm |
| EV2039 | **3.0**/5.5/**0.4**/8.8/**0.5**/8.4/**0.5**/8.3/**0.6**/7.2/**0.7**/6.2/**1.1**/6.5/**2.0**/7.3 | 7.6 kA/cm$^2$<br>24.4 µm |
| EV2091 | **3.0**/5.7/**0.4**/9.1/**0.5**/8.7/**0.5**/8.6/**0.6**/7.5/**0.7**/6.5/**1.1**/6.8/**2.0**/7.6 | 5.7 kA/cm$^2$<br>25.1 µm |
| EV2092 | **3.0**/5.8/**0.4**/9.3/**0.5**/8.9/**0.5**/8.7/**0.6**/7.7/**0.7**/6.6/**1.1**/6.9/**2.0**/7.8 | 5.8 kA/cm$^2$<br>25.3 µm |
| EV2106 | **3.0**/5.5/**0.3**/8.7/**0.4**/8.3/**0.4**/8.2/**0.5**/7.1/**0.6**/6.1/**1.1**/6.6/**2.0**/7.0 | 6.7 kA/cm$^2$<br>26.0 µm |
| EV2107 | **3.0**/5.9/**0.3**/9.3/**0.4**/8.9/**0.4**/8.8/**0.5**/7.7/**0.6**/6.6/**1.1**/7.0/**1.6**/8.0 | 9.0 kA/cm$^2$<br>25.6 µm |
| EV2128 | **3.0**/6.0/**0.3**/9.3/**0.4**/8.9/**0.4**/8.8/**0.4**/7.7/**0.5**/6.3/**0.8**/6.6/**1.4**/7.5 | 10.3 kA/cm$^2$<br>26.3 µm |

- $In_{0.53}Ga_{0.47}As/GaAs_{0.51}Sb_{0.49}$ reference QCL layer structures

| ID | Layer structure | $J_{th}$ & λ<br>(Low Temp) |
|---|---|---|
| EP1561 | **4.8**/5.4/**0.7**/8.6/**0.9**/8.2/**0.9**/8.1/**1.0**/7.1/**1.2**/6.1/**1.6**/6.4/**3.0**/7.2 | 3.7 kA/cm$^2$<br>25.5 µm |
| EP1562 | **4.8**/5.4/**0.6**/8.6/**0.8**/8.2/**0.8**/8.1/**0.9**/7.1/**1.1**/6.1/**1.6**/6.4/**3.0**/7.2 | 5.3 kA/cm$^2$<br>26.6 µm |

**Table S1. Layer structures and properties of all the grown samples.** Threshold current densities ($J_{th}$) and emission wavelengths ($\lambda$) were measured around $T = 50$ K.

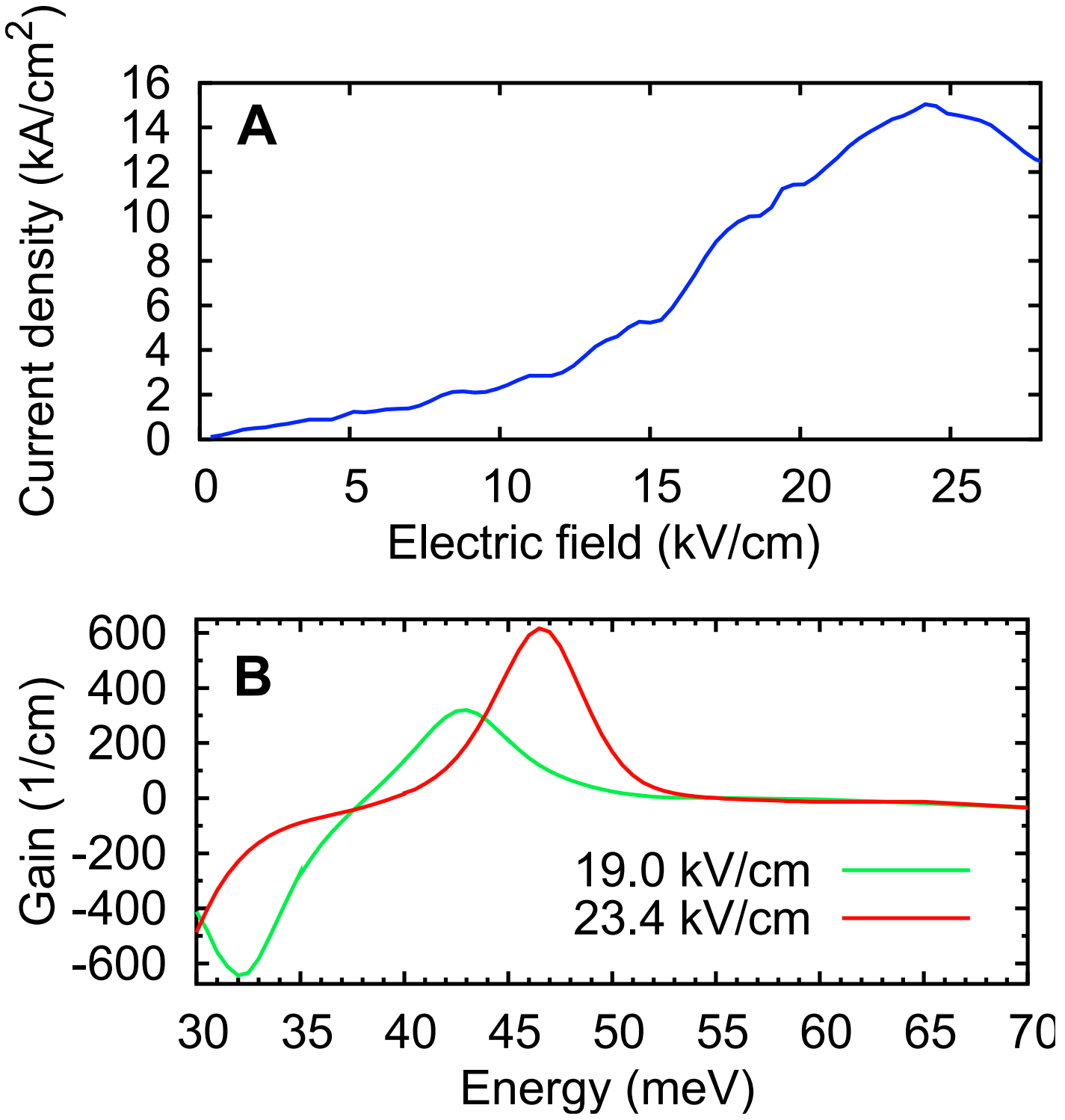

**Figure S3. Non-equilibrium Green's function simulation results.** (**A**) Simulated current density at a lattice temperature of 100 K. The maximum current density is close to the experimental one (13 kA/cm$^2$). (**B**) Simulated optical gain at two applied electric fields; close to the experimental threshold current (green) and close to the maximum current density (red). For electric fields between these two, the gain peak lies in the region of anomalous dispersion between the TO and LO phonon energies. At the lower electric field, the optical gain at an energy of 48 meV is 65/cm, just below the calculated optical losses of 68/cm.